\documentclass[12pt]{article}
\usepackage[text={6.1in,9.1in},centering]{geometry}

\usepackage[applemac]{inputenc}
\usepackage{fontenc}
\usepackage{amsmath}
\usepackage{wasysym}

\usepackage[none]{hyphenat}

\usepackage{epsf,epsfig,graphicx}

\makeindex

\begin{document}

\title{\bf{{\Large{Galileo Galilei and Satellite Navigation}}}}
\author{Alessandro De Angelis}

\graphicspath{{fig/}}



\maketitle

\begin{abstract}
In 1492, for the first time, an unknown ocean opened up before sailors: weeks of navigation and no idea how to pinpoint their location. Since ancient times, navigators had known how to determine latitude by using the North Star, but the ``problem of longitude” was different. More than a century later, Galileo Galilei discovered in Padua Jupiter’s satellites  and quickly realized that a sailor who could observe their eclipses would know his own longitude. Yet his brilliant insight was 400 years ahead of the technology of his time. Impractical at sea, on land this idea became a formidable tool for cartography and ushered in the age of the image of the world. Today the technique can be realized thanks to artificial satellites, and the Tuscan genius’s name has reached space with the European satellite system ``Galileo.”

An exhibition in Paris, organized by the Permanent Representation of Italy to the International Organizations, Sorbonne University, and the Galileo Museum in Florence, and directed by Asia Ruffo di Calabria of the Mus\'ee des Arts et M\'etiers, by Quentin Cheval-Galland of the Sorbonne University, and by Alessandro De Angelis, allowed visitors to observe inventions of the time and some writings by Galileo on the theme of geolocation. The exhibition was held in Paris in June 2024. It was replicated in Prague in October 2024, in Amsterdam in December 2025, and at the Perimeter Institute in Waterloo, Canada, in February 2025.
\end{abstract}

\begin{figure}[h]
\begin{center}\includegraphics[width=.6\linewidth]{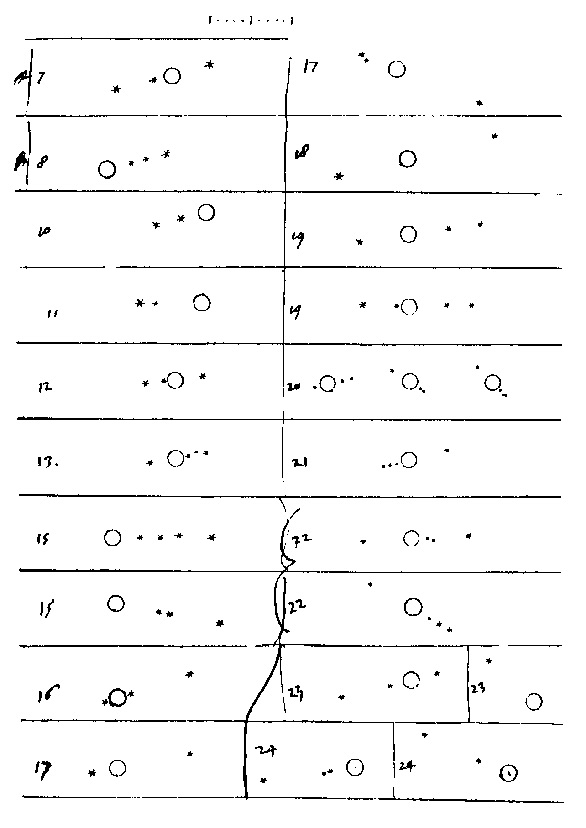}\end{center}
\caption{After observing Jupiter's satellites for the first time from his home in Padua in January 1610, Galileo recorded their positions day by day [page 70r of the Galilean manuscript Gal. 50, Biblioteca Nazionale Centrale, Florence].}
\end{figure}

In 1610, the fortysix-year-old Galileo Galilei (1564-1642), then a professor in Padua, discovered the four large satellites of Jupiter (Figure 1) and observed that they orbit the planet with periods of a few days. He immediately realized that their positions provided an absolute clock that could be consulted from Earth. The uses of this clock were a magnificent obsession that haunted him for over thirty years, until his death. A navigator at sea could compare the local time of an eclipse (on average there are two or three every night) of a Jupiter satellite with the time it was expected to occur at a European reference location; the difference would provide the ship's longitude. Longitude, combined with latitude which could be easily calculated by measuring with a sextant the altitude of the Polaris star or similar references in the southern hemisphere, would allow geolocation. These were the years of great naval explorations: the major world powers vied for the conquest of new colonies, and a discovery like this was worth a fortune. Spain promised tens of millions of dollars at the current exchange rate to anyone who solved the problem. At the end of 1610, Galileo moved to Florence. In 1612, he described his invention to the prime minister and the Florentine ambassador in Madrid, and asked for their help in selling the new technique to the King of Spain. The Spaniards took six years to evaluate the idea and concluded that it was impossible to observe Jupiter's satellites from the deck of a ship at sea due to rolling and pitching, as well as the difficulty of calculating eclipse times.

\begin{figure}
\begin{center}\includegraphics[width=0.5\linewidth]{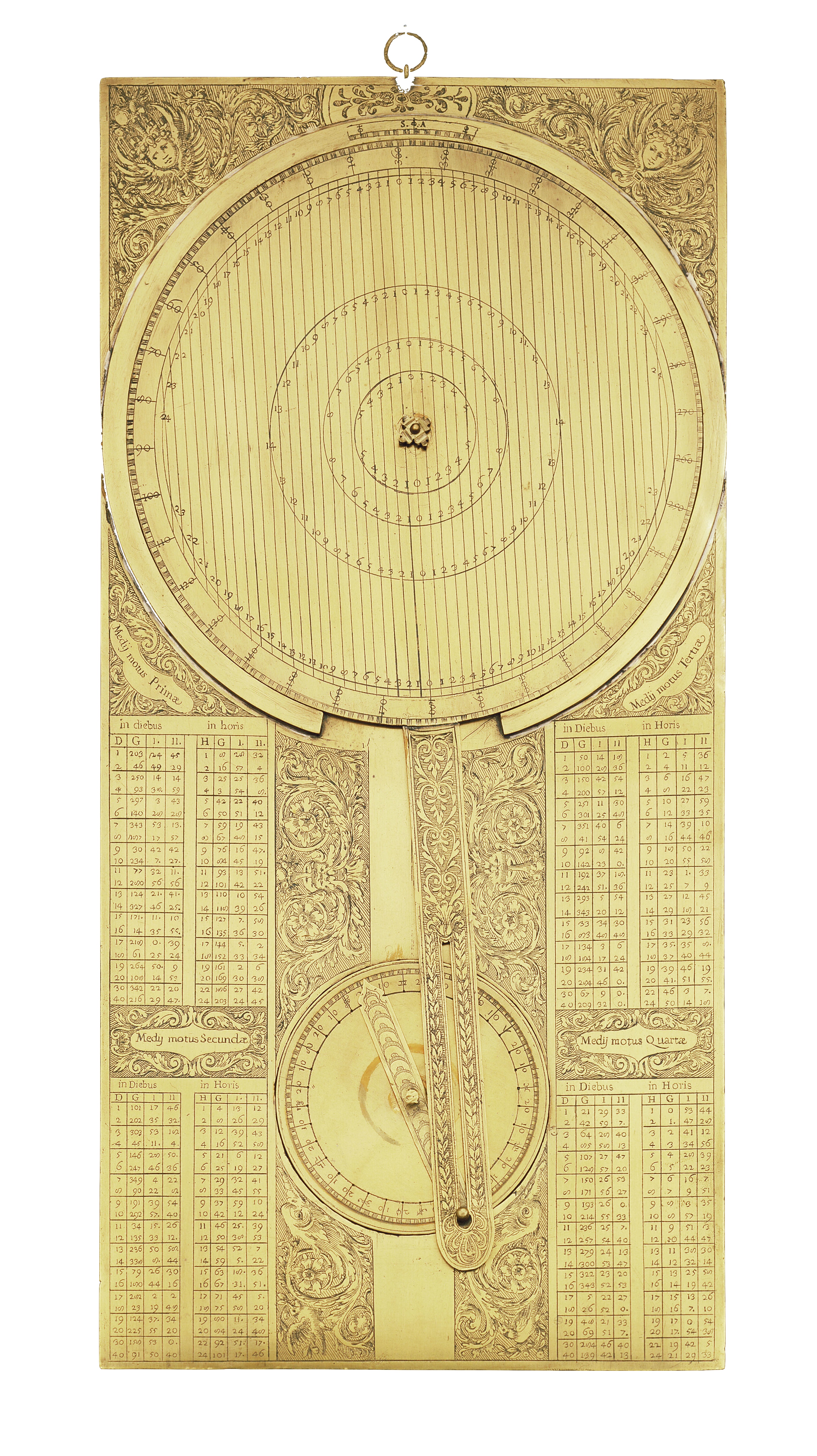}\end{center}
\caption{The ``jovilabe" invented by Galileo to calculate the positions of Jupiter's satellites and the time of their eclipses as seen from Earth. It is an analog calculator similar to an astrolabe, in which the two rotating disks of different diameters, connected by a movable rod, allow the appearances of the satellites observed from Earth to be referred back to the Sun [Museo Galileo, Florence, inv. 3178].} 
\end{figure}

Galileo continued to refine his idea. He invented instruments intended to solve practical problems:
\begin{itemize}
\item The ``jovilabe" (Figure 2), an analog calculator that determined the positions of Jupiter's satellites and was named for its resemblance to an astrolabe. 
\item The ``celatone" (great helmet, Figure 3), a particular setup of the telescope that served to compensate for the ship's vibrations. A telescope was fixed to a headgear at one of the holes corresponding to the eyes. With the other eye free, the observer could spot Jupiter's light in the sky. The celatone allowed viewing the planet's moons by compensating for the ship's movements thanks to the movements of the shoulders and neck. 
\item Lastly, the oleodynamic suspension: a round pool filled with oil on which floated a convex hemisphere; an astronomer sailor was supposed to sit on this hemisphere to dampen the ship's vibrations and conduct observations. 
\end{itemize}
Despite his efforts, Galileo received a final negative opinion from Spain, and he stopped thinking about the matter for a while to focus on his work {\em Dialogue Concerning the Two Chief World Systems,} the most famous of his books - and the one that would cause him the most problems.

\begin{figure}
\begin{center}\includegraphics[width=0.523\linewidth]{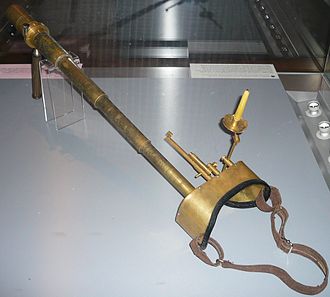}\includegraphics[width=.4\linewidth]{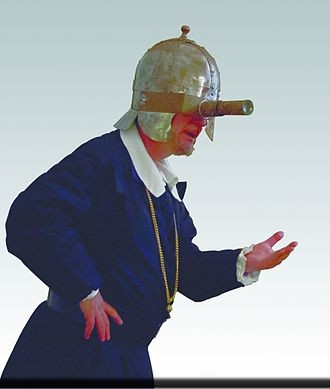}\end{center}
\caption{The ``celatone" (large helmet), which was created by Galileo to observe Jupiter and its moons from the deck of a ship, overcoming problems related to rolling and pitching. The original instrument, which did not work, has not survived. The photo shows two imaginative reconstructions: the one on the left is at the Greenwich Observatory Museum near London, the one on the right at the  Museo Galileo in Florence.}
\end{figure}

After his heresy conviction in 1633 at the age of 69, and after completing the draft of his last book, {\em Discourses and Mathematical Demonstrations Concerning Two New Sciences,} Galileo took up the issue of geolocation again in 1636. This time, he decided to negotiate through intermediaries with the States General of the Netherlands. Despite rewarding Galileo with a precious gold necklace, which the Inquisition forbade him to accept, the Dutch technical committee reached the same conclusion as the Spanish one.

Galileo did not stop working on perfecting his idea until 1642, when he died.

\begin{figure}
\begin{center}\includegraphics[width=0.7\linewidth]{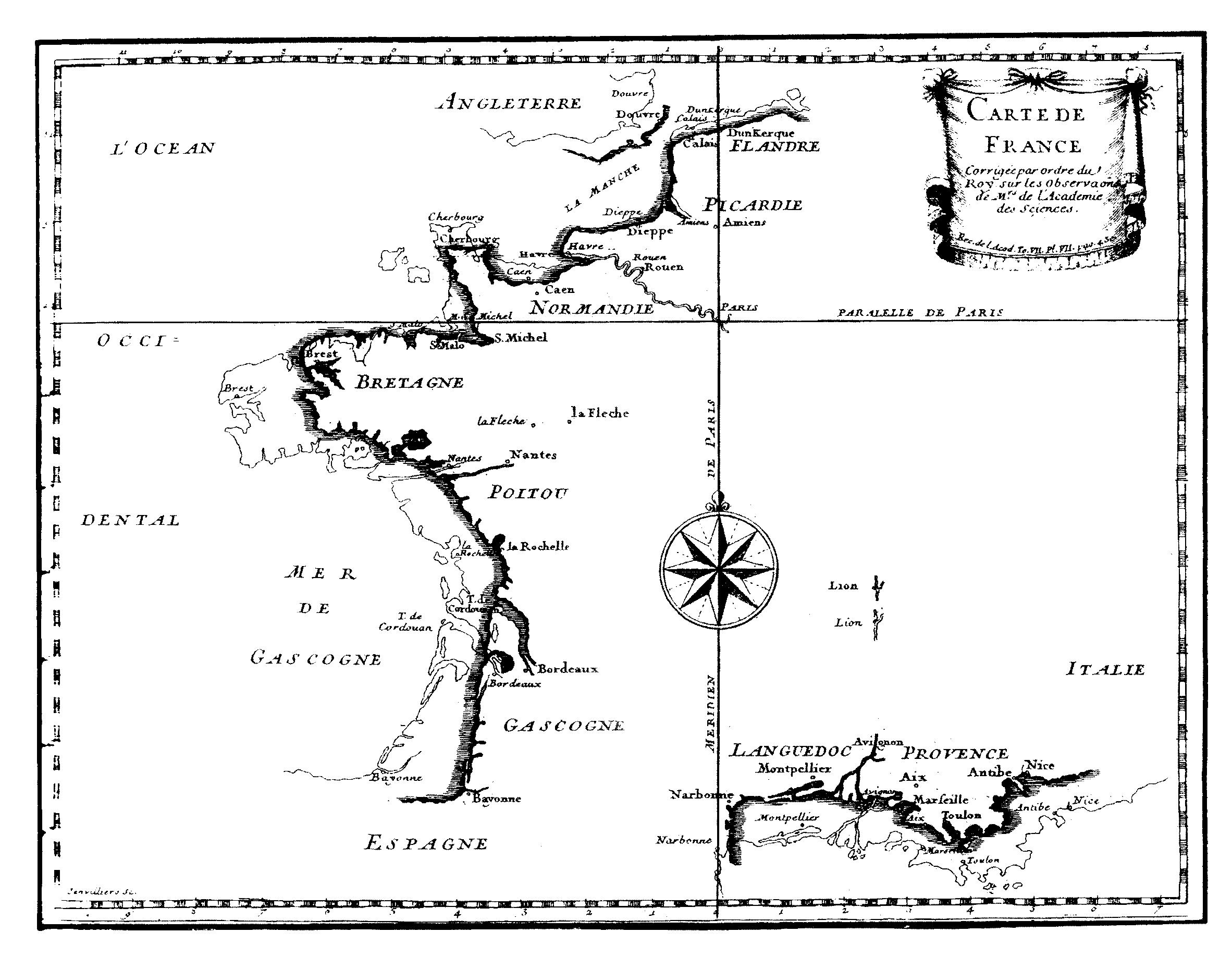}\end{center}
\caption{Map of France corrected in 1682 by Cassini and collaborators using Galileo's geolocation technique [Mémoires de l’Académie royale des Sciences, T7. Biblioth\`eque Nationale de France, Richelieu building].}
\end{figure}

Galileo's method, however, was successfully used, although not from a ship. Giovanni Cassini, a professor in Bologna, in 1671 was chosen as the first director of the Paris Observatory; King Louis XIV, also known as Le Roi Soleil (the Sun King), commissioned an accurate map of France. Cassini started drawing the coasts, for which he used Galileo's method. In 1682, he presented to the Paris Academy of Sciences the first draft of the map (Figure 4). The map shows a comparison with the previous official map. The difference is significant, particularly along the western coasts. It seems that Louis XIV commented that astronomers had stolen more land from him than he had gained in all his wars.

Satellite navigation, technically impossible in Galileo's time due to the excessive distance between Jupiter's satellites,  their relative mutual angular proximity, and the difficulties of pointing the telescopes of that time, became feasible in the space age, using constellations of artificial satellites as references. This fact was somewhat anticipated by Galileo's disciple Vincenzo Viviani, now buried alongside his master in Santa Croce (Florence), who had followed Galileo in the final stages of this long adventure. Viviani wrote in 1564, in his {\em Life of Galileo:} ``this method alone will one day be practiced by all observers on land and sea." Of the four satellite constellations for global geolocation currently in orbit (American, Russian, Chinese, and European respectively), the European one is named after Galileo.


\end{document}